# Electrically pumped quantum-dot lasers grown on 300 mm patterned Si photonic wafers

Chen Shang[1,†], Kaiyin Feng[2,†], Eamonn T. Hughes[1], Andrew Clark[3], Mukul Debnath[3], Rosalyn Koscica[1], Gerald Leake[4], Joshua Herman[4], David Harame[4], Peter Ludewig[5], Yating Wan[2], and John E. Bowers[1,2,*]

[1] *Materials Department, University of California Santa Barbara, Santa Barbara, California 93106, USA*
[2] *Department of Electrical and Computer Engineering, University of California Santa Barbara, Santa Barbara, California 93106, USA*
[3] *IQE, Inc., Greensboro, North Carolina, 27409, USA*
[4] *RF SUNY Polytechnic Institute, Albany, NY 12203, USA*
[5] *NAsP$_{III/V}$ GmbH, Marburg, Germany*

† Authors contributed equally
*Corresponding author: bowers@ece.ucsb.edu

## Abstract

Monolithic integration of quantum dot (QD) gain materials onto Si photonic platforms via direct epitaxial growth is a promising solution for on-chip light sources. Recent developments have demonstrated superior device reliability in blanket hetero-epitaxy of III-V devices on Si at elevated temperatures. Yet, thick, defect management epi designs prevent vertical light coupling from the gain region to the Si-on-Insulator (SOI) waveguides. Here, we demonstrate the first electrically pumped QD lasers grown on a 300 mm patterned (001) Si wafer with a butt-coupled configuration by molecular beam epitaxy (MBE). Unique growth and fabrication challenges imposed by the template architecture have been resolved, contributing to continuous wave lasing to 60 °C and a maximum double-side output power of 126.6 mW at 20 °C with a double-side wall plug efficiency of 8.6%. The potential for robust on-chip laser operation and efficient low-loss light coupling to Si photonic circuits makes this heteroepitaxial integration platform on Si promising for scalable and low-cost mass production.

## Introduction

Internet data traffic has seen a compound annual growth rate of 27% in the last few years and exceeded one zettabyte in 2017. Leveraging the established CMOS processing, Si photonics promises to fulfill this soaring demand for high volume applications such as more efficient, higher capacity, and lower cost interconnects with numerous successes in high performance passive components on 300 mm Si[1–3]. However, due to the indirect bandgap nature of Si and Ge, integrating III-V lasers with Si photonics is a vital step towards widespread adoption. Mass production has been achieved using wafer bonding of III-V gain regions to Si-on-Insulator (SOI) wafers where light is evanescently coupled to the Si waveguide underneath[4–6]. Alternatively, direct growth of III-V gain material onto Si substrates proves to be a more economical favorable solution, which not only eliminates the need for III-V wafers and the complex bonding process, but also possesses additional benefits of compact packaging and better heat sinking. The main challenge is the inevitable crystalline defects that form when merging the III-V and Si crystals via epitaxial growth. To date, tremendous advancements have been made in blanket hetero-epitaxy of III-V compound semiconductors on Si substrates. Anti-phase domains (APDs) from growing polar III-V onto non-polar single step (001) Si have been solved with sophisticated surface preparation[7–14]. Asymmetric step-graded filter was invented such that the threading dislocation density (TDD) has been reduced to as low as $1\times10^6$ cm$^{-2}$ within a 2.55 μm GaAs virtual substrate grown on (001) Si[15]. The coefficient of thermal expansion (CTE) has recently been discovered to be responsible for the formation of misfit dislocations (MDs) near the active region as the film stress state changes from compression to tension during the post-growth cooling process. In lasers with low TDD, the MDs are more detrimental to the device reliability than TDs considering their larger interaction area with the active region[16]. To combat with the MDs, a thin strained quantum wells (QWs) was inserted as trapping layers (TLs) above and below the active region to block the MDs away from the active region[17]. This research field was further boosted by using a quantum dot (QD) active region in place of the traditional quantum wells (QWs). Due to the atom-like density of states, switching to QDs as active not only provides a lower entry to Si photonics thanks to the greatly reduced sensitivity to non-radiative defects, but also possess numerous unique properties that are beneficial to photonic integrated circuits. Great stride has been made in individuals QD devices grown on Si, showing high temperature stability[18], low threshold operation[19,20], and low reflection sensitivity[21], etc. As a result of these defect management advances and innovations in the active region, the extrapolated lifetime of epitaxially grown QD lasers by MBE on CMOS compatible blanket Si device show minimum degradation after more than 4000 hours of constant current stress at 80 °C, with an extrapolated life time exceeding 200,000 hours[22].

However, the thick buffer layers required to reduce the TDD prevent evanescent coupling to the underlying Si waveguides and have hindered the integration of epitaxial QD lasers with conventional Si photonics. One solution is to grow the III-V gain material in pockets and butt-coupled to SiN or SOI waveguides embedded in the surrounding matrix material, as conceptually depicted in Figure 1.

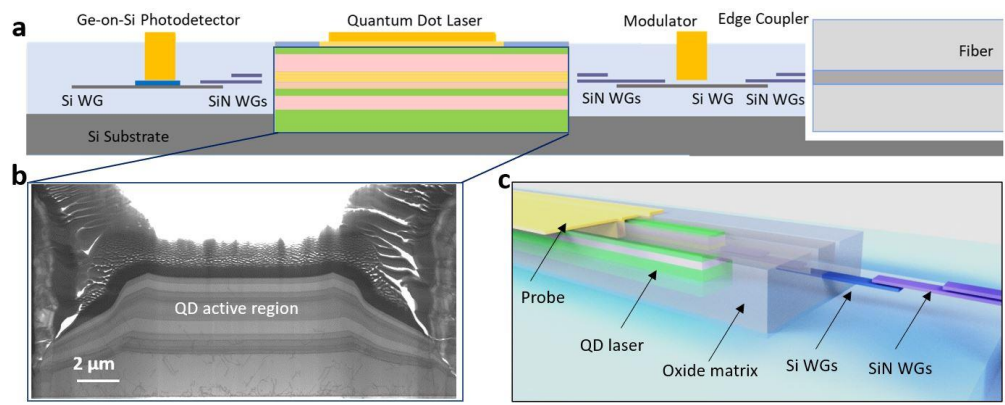

Figure 1: (a) Schematic diagram of the heteroepitaxial laser integration on Si platform. (b) Cross-section TEM of early attempts on growing full laser epi in the pockets. (c) Schematic of intended device configuration.

Passive waveguide coupling[23] or optically pumped nano-ridge lasers on patterned 300 mm Si[24] have previously been demonstrated. The vertical alignment of the QD active region to the waveguides can be precisely controlled during growth. Such in-pocket configuration would also mitigate the issues with the residual tension, possibly yielding lasers with higher reliability compared to those grown on blanket Si[25,26]. However, transferring the epi stack, especially the QD active region, from the blanket Si substrate into the pockets on the patterned Si template is a non-trivial task. In this work, we have identified and solved the additional growth complications induced by the template architecture and have achieved high quality QD nucleation in the pockets. As a proof of concept for this integrated photonic platform with a CMOS process on a 300-mm Si wafer, we demonstrate, to the best of our knowledge, the first electrically pumped in-pocket Fabry-Perot QD lasers emitting around 1300 nm with cleaved facets, sustaining lasing characteristics up to 60 °C with a wall plug efficiency of 8.6%.

**Materials and Methods**

**High quality III-V QD laser material grown in the oxide pockets**

As depicted in Figure 2(a), the initial 4.5 µm $SiO_2$ was deposited by low frequency plasma chemical vapor deposition using tetraethoxysilane (TEOS) on CMOS compactible 300 mm Si wafers. Rectangular pockets were then dry etched down to the Si surface with a pattern fill factor of 95%, followed by a Si recess to make room for the defect-managing buffer layers with InGaAs dislocation filters. To protect the recessed Si sidewall and avoid lateral growth of III-V in the pocket, a thin $SiO_2$ layer was then deposited on the sidewall. A thin 200 nm APD-free GaP layer followed by 500 nm of GaAs was then selectively deposited in a 300 mm MOCVD reactor at $NAsP_{III/V}$ GmbH as a seed layer. The intended buffer structure for the QD laser consisted of a 1-µm GaAs, followed by the InGaAs asymmetric graded dislocation filter layers grown at 500 °C, and completed with an additional 300 nm of GaAs. The active region consisted of 5 layers of InAs QDs, grown at 495 °C, embedded in 7 nm $In_{0.15}Ga_{0.85}As$ QWs with 40 nm GaAs barriers. Surrounding the active region were the AlGaAs cladding regions and the *n*- and *p*-GaAs contact layers. The structure is schematically illustrated in Figure 2(b).

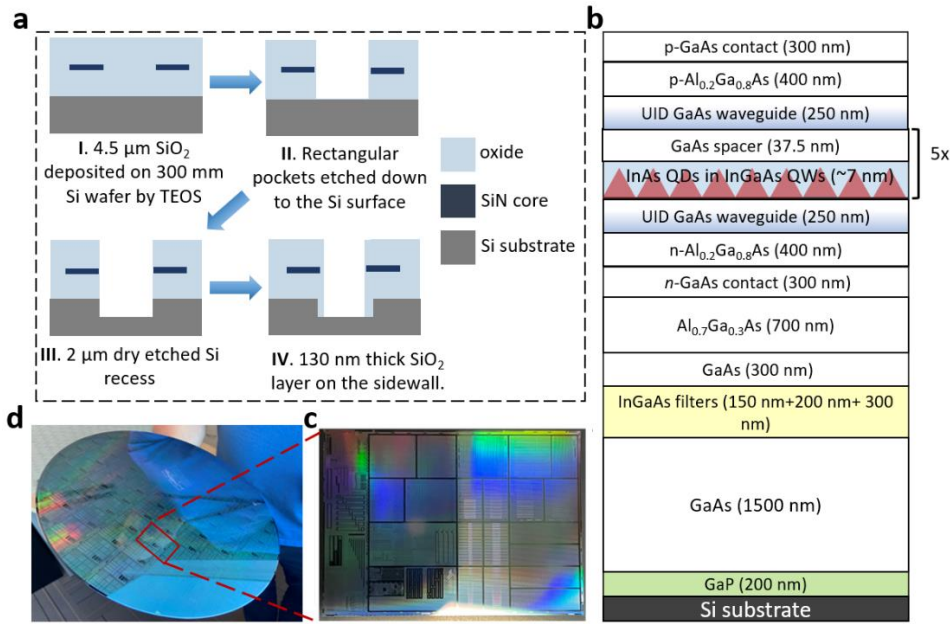

Figure 2: (a) Schematic diagram of the pattern formation process before III-V deposition. (b) Simplified III-V laser stack. (c) the diced coupon from the as-patterned 300 mm Si wafer for growth condition investigation. (d) As-patterned 300 mm Si wafer

Achieving the goal of an electrically pumped QD laser requires a precise understanding of growth condition drifts due to the patterned oxide surrounding the pockets and the resulting III-V epitaxy quality. This work was undertaken using coupons (Figure 2(c)) diced from an equivalent 300 mm Si template (Figure 2(d)) that were transferred to a Veeco Gen-II 3-inch solid source MBE system at UCSB for depositing the same QD laser structure described above. Since the majority of the template area is covered by the oxide mask, using reflective high-energy electron diffraction (RHEED) as the *in-situ* monitor for the surface quality and QD nucleation is not feasible, and the pyrometer readings are less trustworthy. As shown in Figure 3(a), we observed that, for the same sample geometry, the sample with the oxide mask gives a lower pyrometer reading compared to that measured on blanket GaAs/GaP/Si at the same heater power due to the lower emissivity of the oxide. After the oxide mask is covered with polycrystalline III-V from the non-selective growth, the pyrometer reading is higher than that measured on blanket GaAs/GaP/Si. The thicker the polycrystalline layer, the higher the pyrometer reading, suggesting the pyrometer reading depends on the thermal mass of the absorptive polycrystal III-V. Though selective area MBE growth can be realized with no polycrystal deposition on the oxide mask, the required high growth temperatures and low growth rate would degrade the QD nucleation[27]. Thus, trusting pyrometer profiles calibrated from either the Si template or the polycrystalline covered surface before InAs QD deposition would result in either an under- or over-estimate of the actual growth temperature, respectively. Since the optimal temperature window for QD nucleation is small (±2.5°C), this template architecture introduces large temperature uncertainty, making it challenging to achieve high-quality QD nucleation. Therefore, sample heater power calibrated on a blanket GaAs/GaP/Si template was used for the pocket growth instead of the pyrometer, and blanket-substrate-level III-V film quality was achieved with accurate layer thicknesses, shown later in Figure 4(a). The TDD estimated from cross-sectional scanning transmission electron microscopy (STEM) is $1.5\times10^7$ cm$^{-2}$ above the InGaAs dislocation filter layers, achieving a TDD filtering efficiency of 98%.

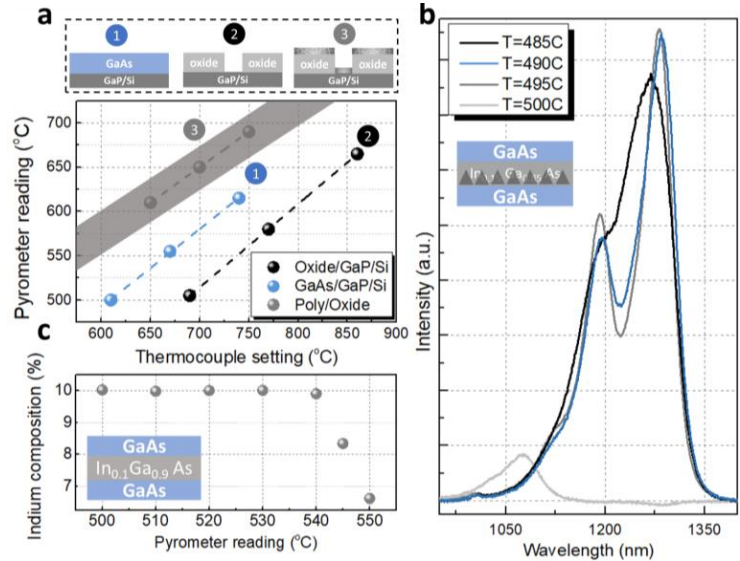

Figure 3: (a) Pyrometer reading discrepancy between different surface conditions. The shaded area indicates the range of readings obtained depending on the polycrystal thickness, with the lower and the upper boundaries indicating the reading measured at the beginning and the end of MBE deposition, respectively. The points in profile 3 represent pyrometer measurements obtained just before InAs QD deposition. (b) PL growth temperature series on native GaAs substrate. (c) Indium composition as a function of growth temperature. The insets of (b) and (c) illustrate the test sample structures.

However, no QD contrast was observed in the cross-section TEM, shown in Figure 4(b, left), and no photoluminescence (PL) signal was obtained from this early structure. Thus, it appears that the same heater power that achieves a surface temperature of 495 °C for QD nucleation on blanket GaAs/GaP/Si, results in a higher temperature in the pockets, potentially evaporating the InAs QD material. We attribute this to more radiation being absorbed by the highly defective polycrystalline III-V from the ambient and less heat being radiated from the mask covered area due to the lower oxide thermal conductivity. To obtain an estimate of the overheating window, separate experiments were conducted on native GaAs substrates. Test structures mimicking the laser active region with one layer of QDs were grown at various pyrometer temperatures. As shown in Figure 3(b), the PL spectrum obtained suggests that an overheating of merely 5 °C above the optimal nucleation temperature would result in severe PL signal degradation due to the evaporated dots. To obtain the upper bound of the template-induced overheating, test structures on GaAs substrates with 50 nm In$_{0.1}$Ga$_{0.9}$As, capped with 10 nm GaAs were grown at different pyrometer temperatures. A loss of indium was observed at a growth temperature above 540 to 545 °C, as shown in Figure 3(c).

A STEM EDS map of the in-pocket InGaAs filter layers was then obtained, showing that the InGaAs filters preserve the designed indium composition, shown in Figure 4(c). This suggests that the overheating from the template architecture is no more than

45 °C. Arbitrary lowering the heater power was then carried out in 5 °C intervals from the calibrated pyrometer profile obtained on blanket GaAs/GaP/Si wafer. High-quality QD nucleation was finally achieved after an approximately 30 °C decrease, shown in Figure 4(b, right). Room temperature PL signal, shown in Figure 4(d), was then obtained from the in-pocket material with a ground state peak near 1300 nm, a full width at half maximum (FWHM) of 32 meV, and a ground-to-excited-state peak separation of 70 meV, comparable to typical values for blanket Si substrates. The above growth conditions form the basis for the process transfer to a 300 mm MBE reactor at IQE, Inc. with similar process trends observed on the larger production platform. The laser results reported here were grown at IQE on 300 mm wafers.

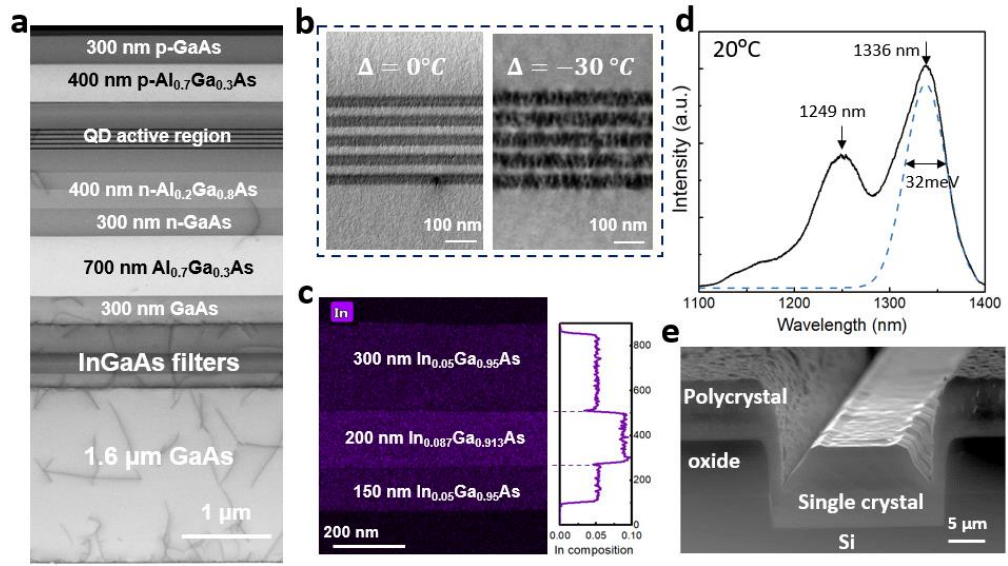

Figure 4: (a) Cross-section TEM of the full in-pocket III-V stack above the 200 nm GaP. (b) Cross-sectional STEM of the active region with different temperature adjustments. Clear QD contrast observed after 30 °C decrease (right). (c) EDS map of the dislocation filter layers labeled with the designed compositions (left) and the extracted actual composition (right). (d) Room temperature PL spectrum from the in-pocket laser material, the dashes line is the Gaussian fit of the ground state line shape. (e) Cross-section SEM of the as-grown in-pocket material.

**Device fabrication and testing**

The as-grown 300 mm wafer, shown in Figure 5(a), was diced into rectangular coupons for processing at UCSB Nanofab facility. The surface morphology presents unique challenges for device fabrication due to the non-selectively deposited polycrystalline III-V on the oxide surface and the trench sidewalls, as well as heavy faceting near the trench edge, previously shown in Figure 1(b) and Figure 4(e). Detailed process development steps were undertaken to transfer processes for blanket-wafer ridge waveguide lasers to fabrication of the in-pocket gain section. Figure 5(b) shows schematics of the fabrication process flow. We developed a wet etch process to selectively remove the polycrystalline III-V on the oxide surface. The removal of surface polycrystal enables easier handling with a more planarized sample surface and better focus calibration in the photolithography steps. Planarized sample surface also serves for better metal liftoff. The alignments were carried out using markers pre-patterned in the SiN waveguide layers on the Si handle wafer. Ridge waveguide was defined with a Cl$_2$-based one-step inductively coupled plasma (ICP) etch that exposes the *n*-doped GaAs contact layer. Thick PECVD SiO$_2$ was used for passivation of waveguide sidewalls as well as pocket edges. We deposited Pd/Ti/Pd/Au for *p*-contact at the ridge-top, and Pd/Ge/Au for *n*-contact on top of the bottom *n*-doped layer and cladding layer respectively. After that, we used dry-etching for via-opening and performed probe metal deposition to form probing pads at wafer surface which are routed towards contacts inside the pockets.

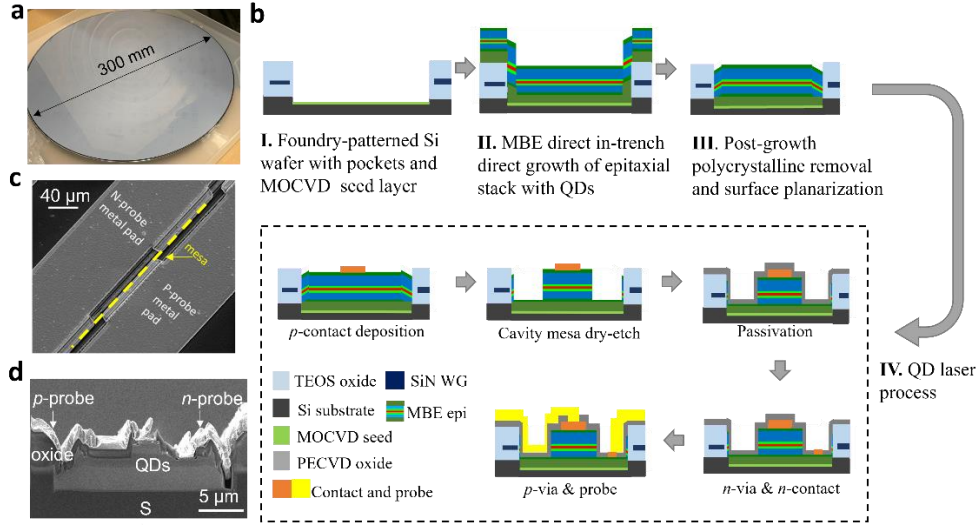

Figure 5: (a) As-grown 300 mm wafer from IQE, covered with polycrystal III-V. (b) Simplified fabrication flow for laser fabrication, not drawn to scale. (c)Top-down view of an as-fabricated device. (d) Cross-section SEM of the as-cleaved in-pocket laser.

Figure 5(c) shows the completed chip with probe metal along the gain-element pockets. To investigate the QD laser quality as a proof of concept, Fabry-Perot cavities were formed with as-cleaved facets along the <110> directions after Si substrate thinning. No further facet coating was applied at this point. Figure 5(d) shows a cross-section SEM of image of a device in a pocket of width around 20 µm. The device has a ridge width of 3.5 µm, and the bright probe-metal layer shows the routing across the pocket-edge step of over 4 µm, demonstrating the feasibility of testing active element inside pocket from wafer surface despite the challenges from the high aspect ratio of the device geometry.

Electrically pumped lasing was achieved under CW conditions with a maximum double-side output power of 126.6 mW at a stage temperature of 20 °C, a threshold current of 47.5 mA, and a series resistance of 3.5 ohms. The highest double-side wall-plug efficiency of 8.6% is achieved at an injection current of 214 mA. Figure 6(a) shows the light-current-voltage (*L-I-V*) of this device. Devices with comparable performance were obtained from different pocket widths on the same bar, such as 20 µm and 30 µm, which indicates relatively high crystal quality uniformity and low geometry-dependent growth rate variation owning to the ballistic nature of MBE growth. Lasers from other process runs and from QDs grown on smaller wafers at UCSB showed similar results. Temperature-dependent measurements of the lasers were also carried out, with *L-I-V* curves of a 2 mm long device shown in Figure 6(b). CW lasing was observed up to 60 °C (pulsed lasing up to 70 °C). Compared to ridge waveguide QD lasers grown on blanket Si wafer, the heat dissipation is relatively limited by the thick oxide layer on the sample surface near the device ridge along the pockets. However, we expect the high-temperature device performance to be improved from further optimization in various aspects such as template quality, growth condition and device design.

To study the lasing wavelength, we aligned a lensed fiber at the cleaved facet to collect the spectra with an optical spectrum analyzer (Yokogawa, AQ6370C). The spectra of a 2 mm long device with 4.5 µm ridge width at various injection currents at a stage temperature of 20 °C are shown in Figure 6(c). Below threshold, the amplified spontaneous emission spectrum shows peak at around 1300 nm, where lasing was achieved beyond threshold with a peak at 1300 nm. A slight red shift in the lasing wavelength has been observed as the injection current increases due to junction heating. The lasing wavelength agrees well with the in-pocket PL obtained after MBE growth prior to fabrication.

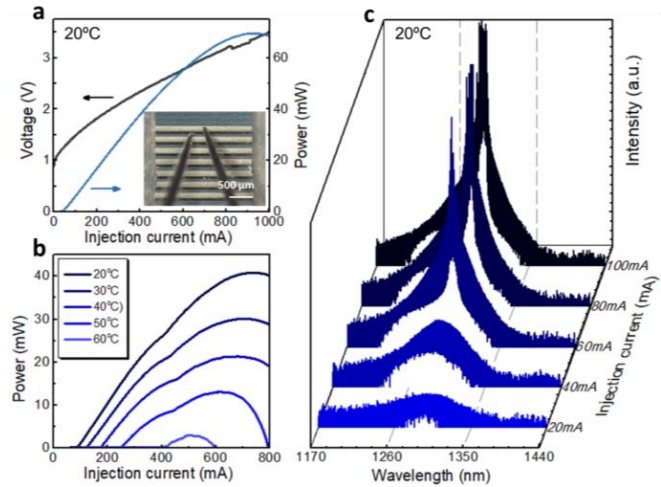

Figure 6: (a) Room-temperature LIV of the device with the highest output power. Insets shows the probe needles on a cleaved laser bar. (b) temperature-dependent LIV showing lasing up to 60 °C. (c) lasing spectrum as a function of injection current at room temperature.

The in-pocket gain element fabrication process can be adapted and transferred to different epitaxial layer stack designs in the AlGaAs/GaAs material system to be used in this on-chip butt-coupled heteroepitaxial integration platform. With additional lithography and etching steps and/or alternative layout designs, a variety of active components such as semiconductor optical amplifiers, mode-locked lasers or distributed-feedback lasers can also be fabricated from the same chip. This fabrication process can be used in integrating complex photonic circuits on 300 mm diameter SOI wafers.

## Discussion

In summary, we present here the first electrically pumped QD lasers epitaxially grown in pockets of patterned 300 mm Si photonic wafers. A maximum double-side output power of 126.6 mW at 20 °C and CW lasing up to 60 °C were observed. Identifying the

MBE growth challenges introduced by the template architecture paves the way for high-quality QD nucleation, with near-blanket-level crystalline quality. It is worth mentioning that the growth conditions, especially the QD nucleation temperature offset, are expected to be dependent on the oxide pattern structure, pattern fill factor, and the polycrystal thickness. Thus, each specific combination of template and epi design would potentially require a specific set of growth conditions and so standardized pocket sizes and pattern designs are desired. Fortunately, the proposed idea for locating the optimal growth window is applicable to other combinations of templates and epi designs. The key challenge would then be reduced to knowing the most temperature sensitive crystalline equality of the specific epi design, including, but no limited to composition and morphology. Selective polycrystal removal neutralizes the potential issue with lithography and metal continuity from the thick non-selective III-V deposition on the oxide mask, resulting in the highly conformal metal routing around the in-pocket laser strip. The QD lasers in this work are a promising approach for monolithically integrated on-chip light sources on CMOS compatible Si wafers.

**Funding.** Defense Advanced Research Projects Agency (No. HR0011-20-C-0142) and Air Force Research Laboratory under AIM Photonics (agreement number FA8650-21-2-1000).

**Disclosure.** The authors have no conflict of interest.

**Data availability.** Data may be obtained from the authors upon reasonable request.

**Acknowledgment**. We thank W. Jin, Z. Zhang, J.C. Norman, M. Dumont, M.J. Kennedy and C. Xiang for useful discussions and help with the fabrication and testing.